\documentclass[fleqn,usenatbib]{mnras}

\usepackage{newtxtext,newtxmath}

\usepackage[T1]{fontenc}

\DeclareRobustCommand{\VAN}[3]{#2}
\let\VANthebibliography\thebibliography
\def\thebibliography{\DeclareRobustCommand{\VAN}[3]{##3}\VANthebibliography}

\usepackage{graphicx}	
\usepackage{amsmath}	

\usepackage[usenames,dvipsnames]{color}

\newcommand{\code}[1]{\texttt{#1}}

\usepackage{placeins}

\title[Core convection in supermassive stars]{3D hydrodynamics simulations of core convection in supermassive main-sequence stars}

\author[S. Blouin et al.]{Simon Blouin$^{1,\dagger}$\thanks{E-mail: sblouin@uvic.ca},
Huaqing Mao$^{2,\dagger}$,
Tyrone E. Woods$^{3}$,
Pavel Denissenkov$^{1,\dagger}$,
Paul R. Woodward$^{2,\dagger}$,
and
\newauthor
Falk Herwig$^{1,\dagger}$
\\
$^{1}$Department of Physics and Astronomy, University of Victoria, Victoria, BC V8W 2Y2, Canada\\
$^{2}$LCSE and Department of Astronomy, University of Minnesota, Minneapolis, MN 55455, USA\\
$^{3}$National Research Council of Canada, Herzberg Astronomy \& Astrophysics Research Centre, Victoria, BC V9E 2E7, Canada\\
$^{\dagger}$Joint Institute for Nuclear Astrophysics -- Center for the Evolution of the Elements (JINA--CEE)
}

\date{Submitted: 18 November 2022, Accepted: 16 March 2023}

\pubyear{2022}

\begin{document}
\label{firstpage}
\pagerange{\pageref{firstpage}--\pageref{lastpage}}
\maketitle
\begin{abstract}
Supermassive stars are Population III stars with masses exceeding $10^4\,M_{\odot}$ that could be the progenitors of the first supermassive black holes. Their interiors are in a regime where radiation pressure dominates the equation of state. In this work, we use the explicit gas dynamics code \code{PPMstar} to simulate the hydrogen-burning core of a $10^4\,M_{\odot}$ supermassive main-sequence star. These are the first 3D hydrodynamics simulations of core convection in supermassive stars. We perform a series of ten simulations at different heating rates and on Cartesian grids with resolutions of $768^3$, $1152^3$ and $1728^3$. We examine different properties of the convective flow, including its large-scale morphology, its velocity spectrum and its mixing properties. We conclude that the radiation pressure-dominated nature of the interior does not noticeably affect the behaviour of convection compared to the case of core convection in a massive main-sequence star where gas pressure dominates. Our simulations also offer support for the use of mixing-length theory in 1D models of supermassive stars.
\end{abstract}


\section{Introduction}
\label{sec:intro}
Over the last decade, several very massive ($\gtrsim 10^9\,M_{\odot}$) quasars at redshift $z \sim 7$ have been identified \citep[e.g.,][]{mortlock2011,wu2015,banados2018}. The existence of such supermassive black holes when the universe was less than 1\,Gyr old is perplexing. Producing a $\gtrsim 10^9\,M_{\odot}$ black hole from a standard $\lesssim 100\,M_{\odot}$ Population~III star by that time would require larger accretion rates than what can be sustained \citep{park2011,whalen2012}. 

To solve this conundrum, the formation of much more massive black hole ``seeds'' is thought to be required. One promising scenario is the collapse of primordial supermassive stars (SMS), Population~III stars with masses $\gtrsim 10^4\,M_{\odot}$ \citep{rees1984,woods2019}. Such stars could be formed from a primordial halo at $z \sim 10-20$ if a strong Lyman--Werner ultraviolet field destroys molecular hydrogen, thereby delaying the collapse of the cloud and preventing its fragmentation into conventional-mass stars \citep{agarwal2012,dijkstra2014,regan2017}. Eventually, the cloud cools from atomic hydrogen line transitions and collapses with infall rates reaching $0.01-1\,M_{\odot}\,{\rm yr}^{-1}$ \citep{latif2013}, thus permitting the formation of SMS. Alternatively, a SMS may be formed in a halo where cold flows drive violent turbulence that prevents star formation until a critical mass is reached and the halo collapses catastrophically \citep{latif2022}. Thanks to the very large infrared luminosities of SMS, the James Webb Space Telescope may soon enable their detection \citep{surace2018,surace2019,whalen2020,woods2021b}.

Modelling the evolution of SMS has been a subject of intense theoretical efforts over the last years \citep{begelman2010,hosokawa2012,hosokawa2013,umeda2016,woods2017,woods2021,haemmerle2018,haemmerle2018b,nagele2020,herrington2023}. There are significant disagreements between independent evolutionary models \citep[e.g., see the review by][]{woods2019}, and the absence of constraining observational data prevents their empirical validation. In this work, we examine more closely one uncertain aspect of the modelling of SMS: the treatment of convection in their cores.

In the hydrogen-burning cores of main-sequence SMS, the total pressure is overwhelmingly dominated by the radiative pressure. More precisely, $\beta \equiv P_{\rm gas}/P \lesssim 0.1$, where the total pressure $P$ is given by
\begin{equation}
    P = P_{\rm gas} + P_{\rm rad} = \frac{R \rho T}{\mu} + \frac{a T^4}{3},
    \label{eq:eos}
\end{equation}
with $R$ the ideal gas constant, $\rho$ the mass density, $T$ the temperature, $\mu$ the mean molecular weight, and $a$ the radiation density constant. Those are rather exotic conditions where the mixing-length theory \citep[MLT,][]{cox1968}, so far used in all evolutionary calculations of SMS, has seldom been tested. The only exception we are aware of is the iron opacity peak convection zone of massive main-sequence stars, where $\beta$ is also of order 0.1 and 3D hydrodynamics simulations have been carried out \citep{jiang2015,jiang2017,schultz2022}.

Here we present the first 3D hydrodynamics simulations of core convection in SMS. Our approach is described in Section~\ref{sec:methods}, where we detail the 1D model used to initialize the 3D simulations and briefly describe the \code{PPMstar} gas dynamics code used in this work. We then present the main results of our simulations in Section~\ref{sec:results} and an MLT analysis in Section~\ref{sec:MLT} before giving our conclusions in Section~\ref{sec:conclusion}.

\section{Methods}
\label{sec:methods}
\subsection{1D models}
To set the initial conditions for our 3D hydrodynamics simulations, we draw from the \code{Kepler} 1D models of \cite{woods2017}. \code{Kepler} is a Lagrangian hydrodynamics and stellar evolution code which includes convective mixing using MLT and uses an adaptive nuclear reaction network coupled to the hydrodynamics \citep[for more details, see][]{weaver1978,woosley2004}. These models are each evolved under a constant accretion rate until the stars either undergo collapse via the post-Newtonian general relativistic instability or approach the end of their nuclear-burning lifetime. They are initialized as $10\,M_{\odot}$ polytropes with central density $\rho_{\rm{c}} = 10^{-3}\,{\rm g\,cm}^{-3}$ and central temperature $T_{\rm c} = 1.2 \times 10^{6}\,$K. The initial protostar is assumed to be chemically homogenous and both its initial composition and that of all accreted material are assumed to be primordial \citep[with abundances following][]{cyburt2001,cyburt2002}. Here we consider a model accreting $0.1\,M_{\odot}/{\rm yr}$ after it has reached a total mass of $\simeq 10{,}000\,M_{\odot}$ as the initial setup for our 3D simulations. This corresponds to a star that is on the main sequence but still early in its evolution.

\begin{figure}
	\includegraphics[width=\columnwidth]{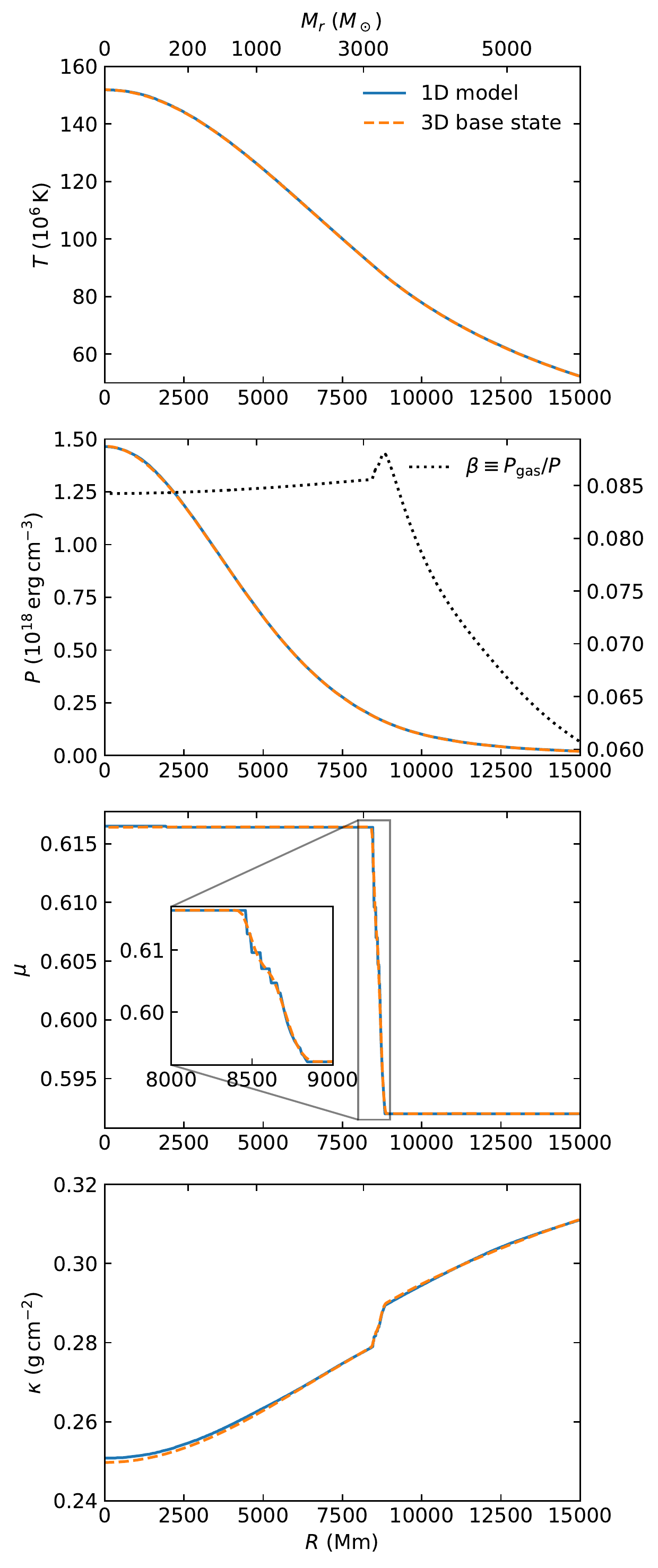}
    \caption{Comparison between the initial 1D \code{Kepler} model (solid blue line) and its representation in \code{PPMstar} (dashed orange line). The hydrogen-burning convective core occupies the $R \lesssim 8400\,{\rm Mm}$ region. The original 1D stratification is accurately recovered in the 3D setup and small-scale numerical noise has been suppressed using spline interpolations (see the $\mu$ profile in the convective boundary region). The second panel also displays the $\beta$ profile (dotted black line, right vertical axis).}
    \label{fig:setup}
\end{figure}

To initialize the 3D simulations, we use the 1D model's central pressure, its entropy ($S$) profile and its $\mu$ profile. From those quantities, a 3D base state is generated by integrating the hydrostatic equilibrium equation and using the equation of state \eqref{eq:eos} implemented in our 3D gas dynamics code. This guarantees the generation of a 3D base state that is exactly in hydrostatic equilibrium. As in our previous works \citep{blouin2022,herwig2022}, small-scale numerical noise is removed from the \code{Kepler} $S$ and $\mu$ profiles using spline interpolations, and a constant entropy is imposed in the convective core. For reference, Figure~\ref{fig:setup} compares the original 1D model to its representation in \code{PPMstar}, after the smoothing procedure. Since we are mostly interested in the behaviour of convection in the hydrogen-burning core, our initial setup only includes layers inside a radius $R_{\rm max}=15{\,}000\,{\rm Mm}$. Most of the radial extent of the radiative envelope is therefore not included, but this inner $15{\,}000\,{\rm Mm}$ nevertheless contains more than half of the mass of the star, $M(R_{\rm max}) \simeq 5600\,M_{\odot}$.

\subsection{\code{PPMstar} simulations}
\label{sec:simulations}
We use the \code{PPMstar} explicit gas dynamics code \citep{woodward2015,jones2017,herwig2022}. Two fluids are included, one with $\mu=0.5920$ representing the envelope material and one with $\mu=0.6164$ representing the heavier core material. \code{PPMstar} now takes into account radiation pressure in its equation of state \citep{mao2022}, a necessary upgrade to simulate SMS. Radiation diffusion is also considered, which is done by including a radiative flux term ${\bf{F}}_{\rm rad} = - \frac{4 a c T^3}{3 \kappa \rho} {\bf \nabla} T$. Direct interpolation of Rosseland mean opacity tables is not practical due to the large computational cost that such a procedure would entail in the context of a highly optimized gas dynamics code like \code{PPMstar}. Instead, we build a simple polynomial fit to the OPAL opacity tables \citep{iglesias1996} in the composition--density--temperature space around the parameters relevant to our simulation setup. As shown in Figure~\ref{fig:setup}, this procedure satisfactorily recovers the opacity profile of the reference 1D model. 

The gravitational acceleration profile is fixed throughout the simulations, meaning that a dynamical collapse of the type expected for a polytrope with $\gamma \leq 4/3$ cannot take place. Convection in the core is driven by heating the central region of the star. Heat is injected in the simulation following a Gaussian profile centered on $R=0$ and with a half width at half maximum of 2160\,Mm, closely matching the 1D model predictions. In order to limit the computational cost of simulating low-Mach number flows, all simulations presented in this work are driven by a heating luminosity $L$ that is $\geq 10$ times the nominal nuclear luminosity $L_{\star}$ of the initial 1D model\footnote{$L_{\star} = 1.485\times 10^{42}\,{\rm erg}\,{\rm s}^{-1}$}, thereby increasing the flow velocity. The properties of the real star can then be estimated by extrapolating from those higher luminosities to the lower nominal luminosity \citep[e.g.,][]{jones2017,herwig2022}. Note that for a simulation with $L= K L_{\star}$, the opacity is set to be $\kappa = \kappa_{\star}/K$, where $\kappa_{\star}$ is the nominal opacity from OPAL. This proportional decrease ensures energy conservation (the additional energy injected in the star can flow through the radiative layers more easily).

Our simulations are performed on Cartesian grids of $768^3$, $1152^3$, and $1728^3$ and they each run for several thousand hours of star time. All runs analysed in this work are listed in Table~\ref{tab:runs}. The different grid resolutions allow to characterize the numerical convergence of our simulations (Section~\ref{sec:rprof}), and the various heating luminosities enable the establishment of the scaling laws required to extrapolate the simulation results to the nominal luminosity (Section~\ref{sec:heating_series}).

\begin{table}
	\centering
	\caption{Summary of the simulations presented in this paper.}
	\label{tab:runs}
	\begin{tabular}{llcc} 
		\hline
		ID & $L/L_\star$ & grid & $t$ (h) \\
		\hline
	    V2 & $10^3$ & $768^3$ & 2880 \\
	    V3 & $10^3$ & $1152^3$ & 2456 \\
	    V4 & $10^2$ & $1152^3$ & 3756 \\
	    V5 & $10$ & $1152^3$ & 5252 \\
	    V7 & $10^2$ & $1728^3$ & 2528 \\
	    V8 & $10^2$ & $768^3$ & 3691 \\
	    V9 & $10^4$ & $768^3$ & 2028 \\
	    V10 & $10^{3.5}$ & $768^3$ & 1831 \\
	    V11 & $10^{2.5}$ & $768^3$ & 1893 \\
	    V12 & $10$ & $768^3$ & 6153 \\
        \hline
		\end{tabular}
\end{table}

\begin{figure}
	\includegraphics[width=\columnwidth]{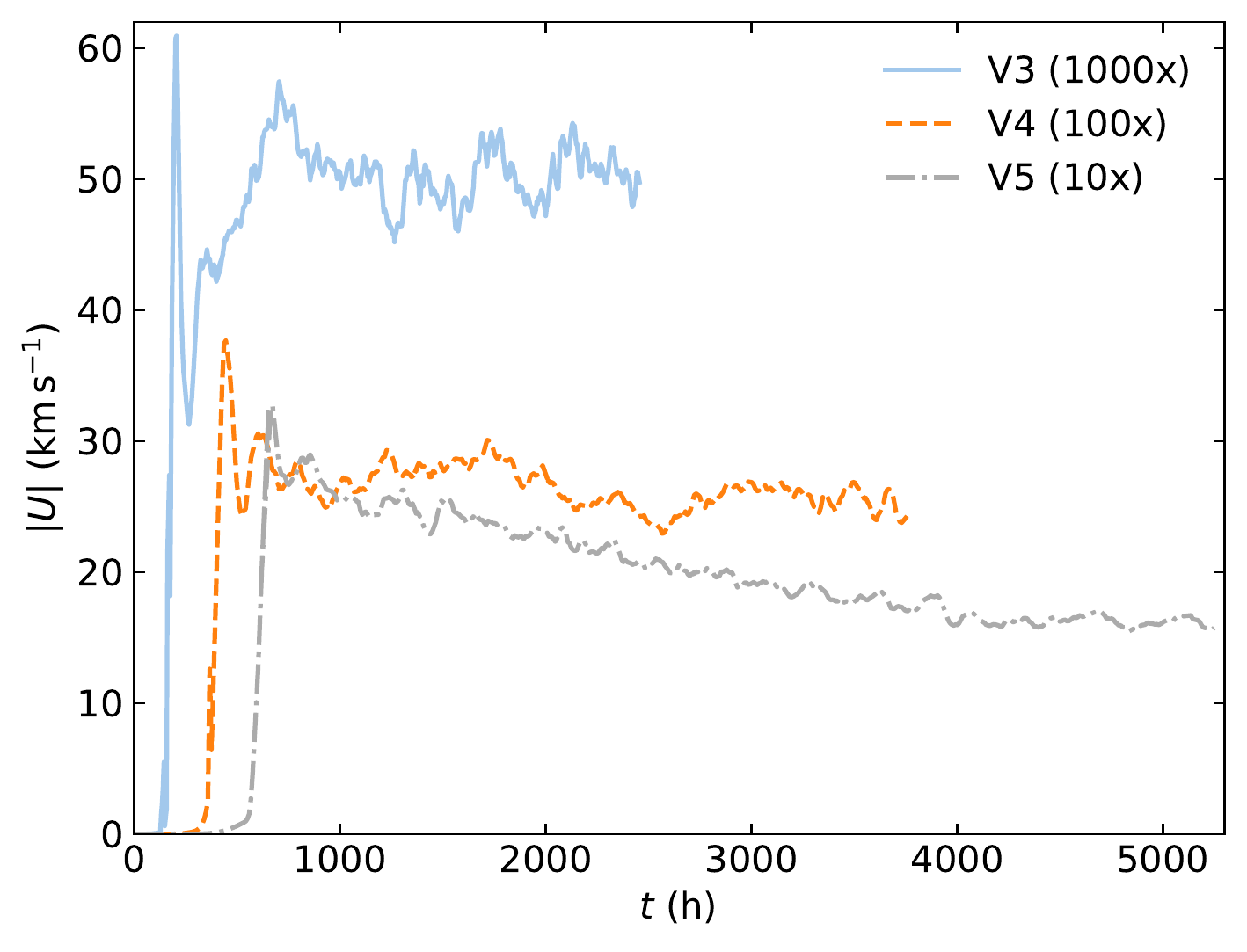}
    \caption{Time evolution of the spherically averaged rms velocity $1\,H_P (\simeq 2400\,{\rm Mm})$ below the convective boundary for runs V3, V4 and V5.}
    \label{fig:convergence}
\end{figure}

Except for V12, which we discuss in more details in Section~\ref{sec:heating_series}, all simulations have run long enough to attain a state of dynamical equilibrium (i.e., the properties of their convective cores reach a steady state). This is shown in Figure~\ref{fig:convergence}, where the spherically averaged rms velocity one pressure scale height below the convective boundary is plotted as a function of time for our three $1152^3$ simulations with different heating luminosities. We can see that V3 reaches dynamical equilibrium after $\simeq 1000\,{\rm h}$. The fact that the lower-$L$ runs require more time to achieve dynamical equilibrium simply reflects their slower dynamics. In the following sections, the initial transient phase is discarded and only the steady-state portion of each simulation is used in our analysis.

It is useful to compare the total simulation times to the convective turnover timescale. We can estimate the convective turnover timescale by taking the diameter of the convective core and dividing it by the rms velocity $|U|$. For run V3 ($L=1000 L_{\star}$), this yields $\tau_{\rm conv} \simeq 17{\,}000\,{\rm Mm}/0.06\,{\rm Mm/s} \simeq 79\,{\rm h}$, meaning that this simulation lasted for about 31 turnover timescales, including $\simeq 20$ past the point where it reaches dynamical equilibrium. This is sufficient to calculate robust mean flow properties.

\section{Results}
\label{sec:results}
\subsection{Center-plane slice renderings}
\label{sec:renderings}
\begin{figure*}
    \begin{center}
    \hspace{-1.2mm}
	\includegraphics[width=88mm]{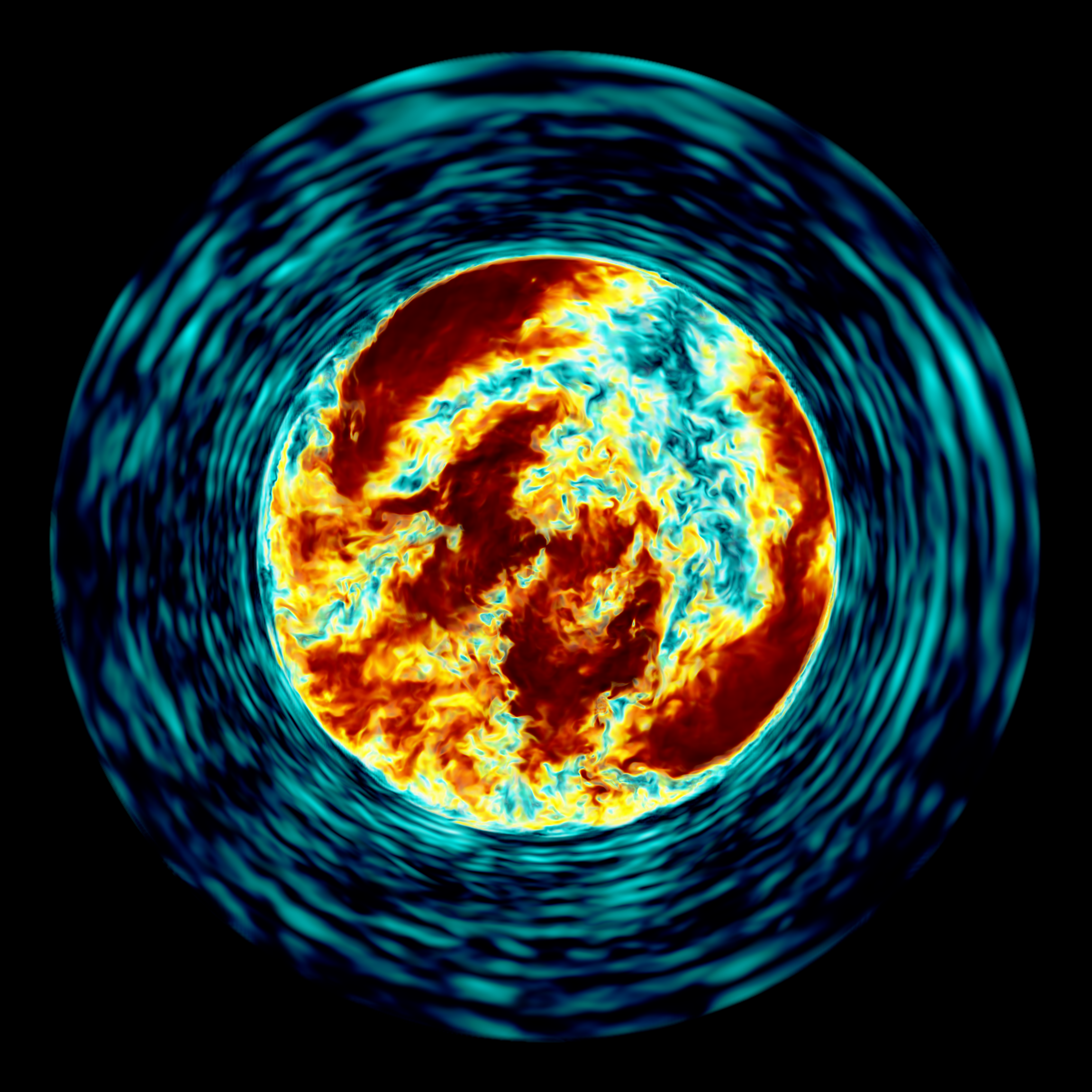}
	\includegraphics[width=88mm]{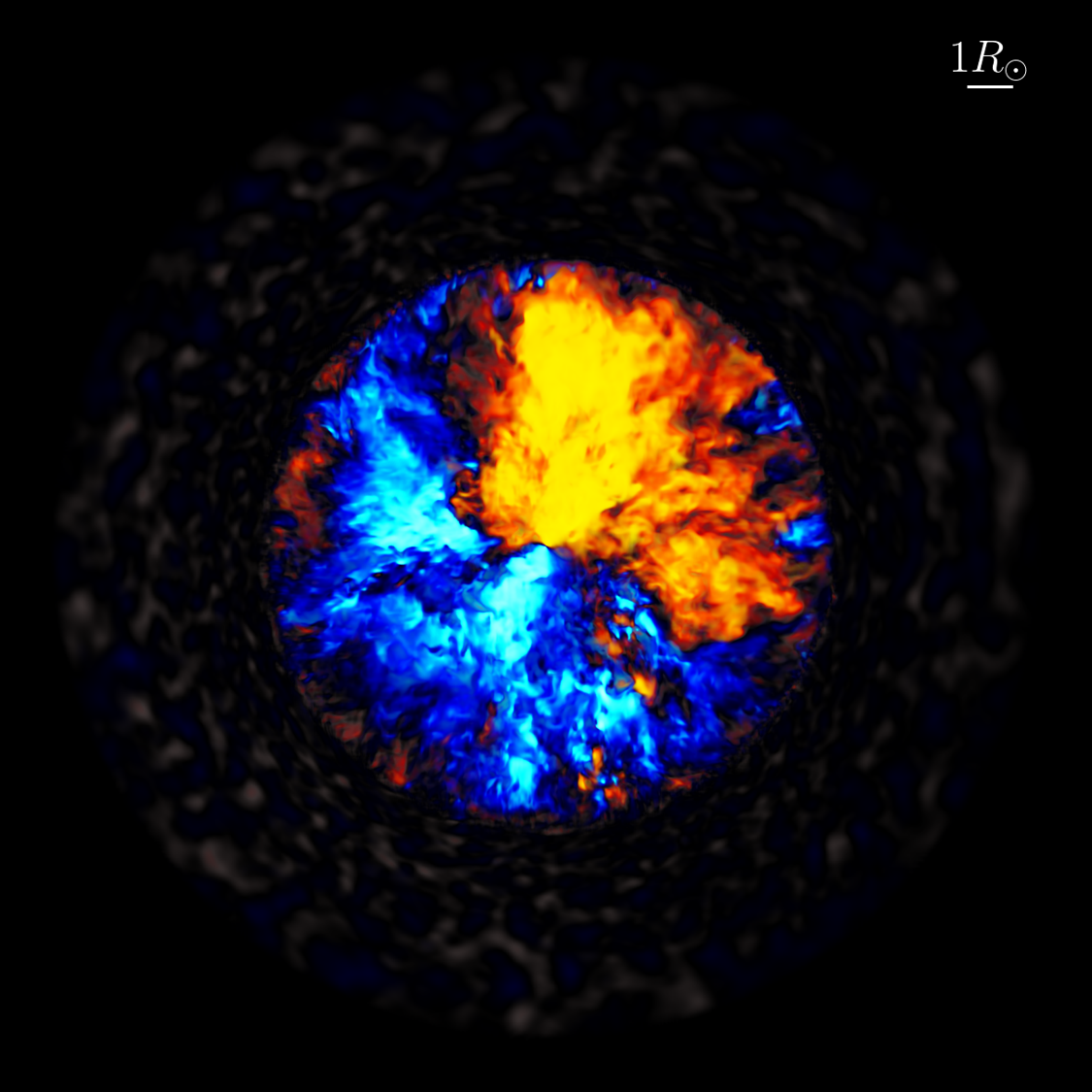}
	\includegraphics[width=88mm]{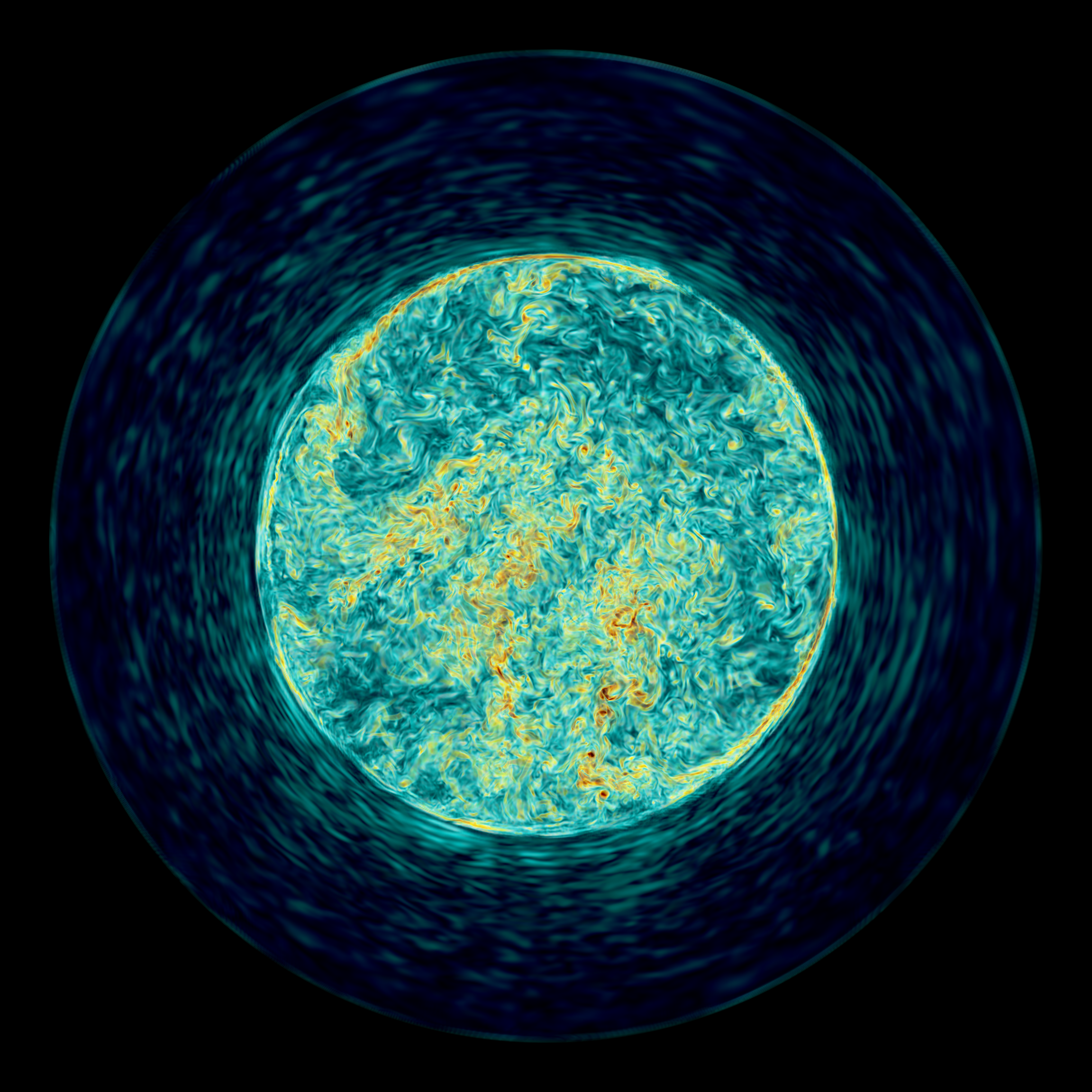}
	\includegraphics[width=88mm]{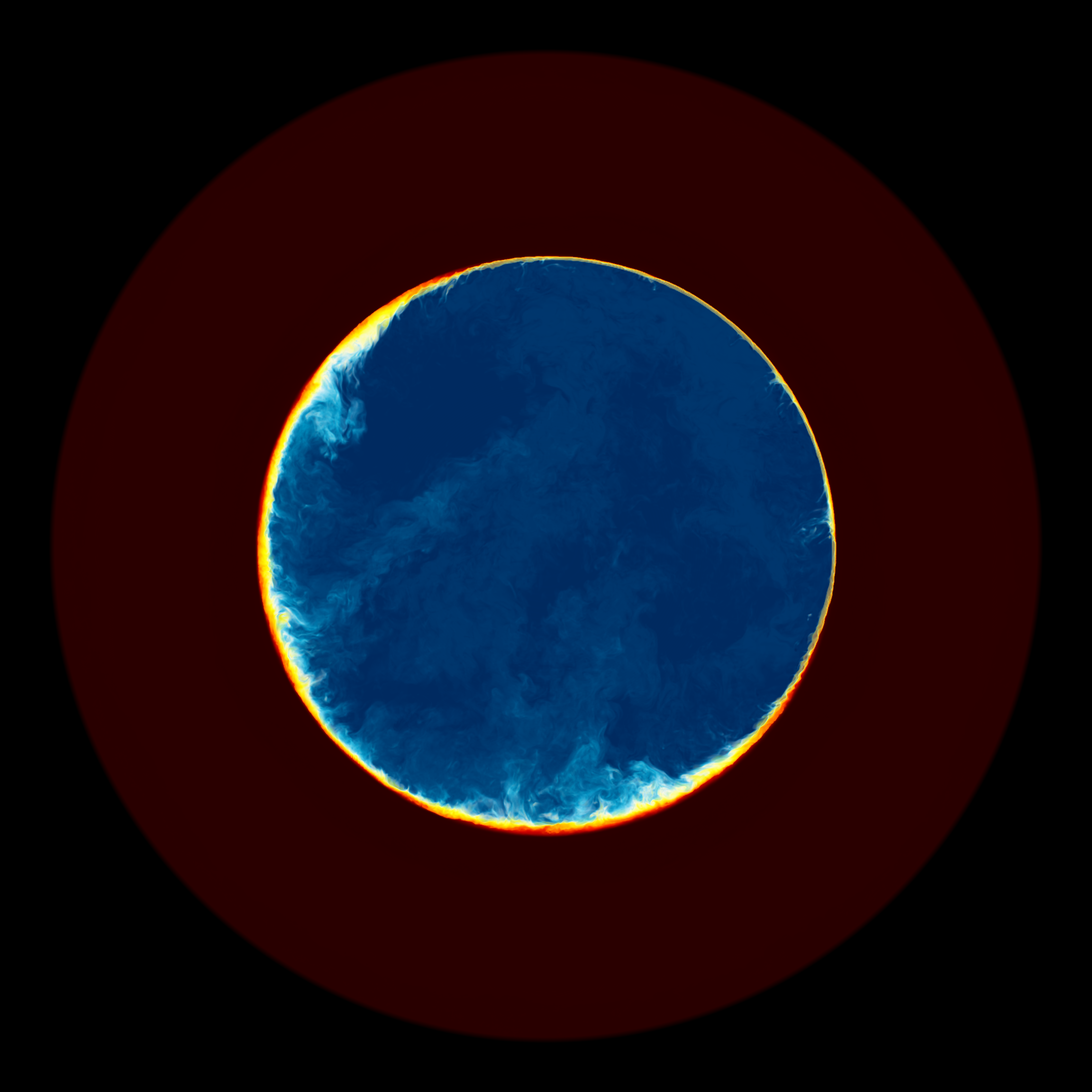}
    \caption{Center-plane slice renderings of run V3 ($L=1000 L_{\star}$, $1152^3$ grid) at dump~626 of the simulation (corresponding to $t=1365\,$h). {\it Top left}: magnitude of the tangential velocity component $|U_t|$, with dark blue, turquoise, yellow, red, and dark red representing a sequence of increasing velocities. {\it Top right}: radial velocity $U_r$, with blue colours representing inward-moving flows (light blue being faster flows than dark blue) and orange colours outward-moving flows (light orange being faster flows than dark orange). {\it Bottom left}: vorticity magnitude, with the same colour sequence as for $|U_t|$. {\it Bottom right}: fractional volume of the envelope fluid ${\rm FV}_{\rm env}$ with dark red being pure envelope material (smaller $\mu$) and dark blue pure core material (larger $\mu$). High-resolution movies are available at \url{https://www.ppmstar.org/}.}
    \label{fig:bobs}
    \end{center}
\end{figure*}

Figure~\ref{fig:bobs} shows center-plane slice renderings of the amplitude of the tangential velocity component $|U_t|$ (i.e., the velocity component perpendicular to the radial direction), the radial velocity $U_r$, the vorticity magnitude $| \nabla \times U|$, and the fractional volume of the envelope fluid (${\rm FV}_{\rm env}$)\footnote{${\rm FV}_{\rm env}$ is the variable used to track the concentration of the envelope fluid: ${\rm FV}_{\rm env}= X_{\rm env} \rho / \rho_{\rm env}$, with $X_{\rm env}$ the mass fraction of the envelope fluid, $\rho_{\rm env}$ its density, and $\rho$ the density of the two-fluid mixture.}. In the first three panels, we can easily distinguish the convective core, characterized by high flow velocities and turbulent motions. The $|U_t|$ and $U_r$ panels clearly show how the flow is dominated by a large dipolar structure. From the center, the material is carried by fast upflows toward the convective boundary in a northeastern direction (in orange in the $U_r$ panel). Upon reaching the core boundary, the flow is forced to turn and then travels mainly in the horizontal direction along the inner contour of the convective core (dark red regions in the $|U_t|$ panel), before eventually separating from the boundary and turning back toward the center (in blue in the $U_r$ panel). This separation is due to the opposing pressure generated by the opposite tangential flow. Facing this pressure gradient and constrained by the convective boundary, the flow is forced to turn inward \citep{herwig2022}. As described in \cite{woodward2015}, this separation generate instabilities that promote the ingestion of envelope material into the core (see the ${\rm FV}_{\rm env}$ plumes traveling inward from the convective boundary in the fourth panel).

This overall flow morphology, dominated by a large dipole structure that goes through the center of the star, is indistinguishable from that observed in our recent \code{PPMstar} simulations of core convection in $25\,M_{\odot}$ main-sequence stars \citep{herwig2022}. This is a first indication that the $P_{\rm rad}$-dominated equation of state that describes the hydrogen-burning cores of SMS has little impact on convection compared to more conventional stellar interiors where $P_{\rm gas}$ dominates.

Finally, note that the ring-like structures clearly visible in the radiative region of the $|U_t|$ and $| \nabla \times U |$ panels are internal gravity waves oscillating in the stable envelope after being excited by the pummeling of the convective boundary. We can infer their nature based on the presence of discrete modes that have frequencies below the local Brunt--V\"ais\"al\"a frequency, as expected for internal gravity waves (see Appendix~\ref{sec:kw}).

\subsection{Radial profiles}
\label{sec:rprof}
\begin{figure}
	\includegraphics[width=\columnwidth]{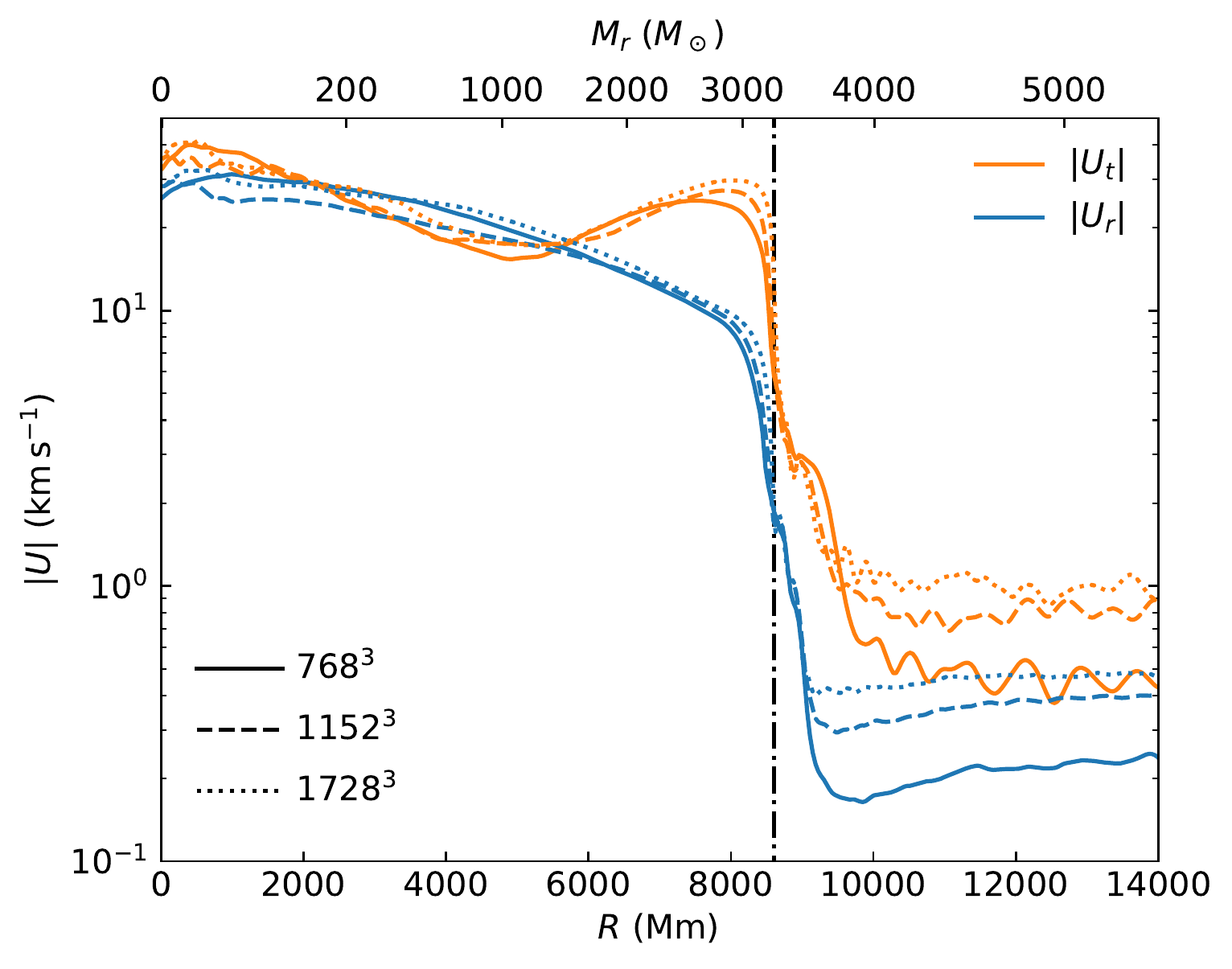}
    \caption{Spherically averaged radial (blue) and tangential (orange) velocity components at $t=2500\,{\rm h}$ for runs V8 ($768^3$ grid, solid lines), V4 ($1152^3$, dashed lines) and V7 ($1728^3$, dotted lines). The vertical dash-dotted line marks the location of the convective boundary, identified by finding the location of the minimum $U_t$ gradient \citep{jones2017}. Note that the simulations shown here were driven with $100\times$ the nominal luminosity (see Table~\ref{tab:runs}).}
    \label{fig:vel_rprof}
\end{figure}

Spherically averaged radial profiles of the rms radial and tangential velocity components are shown in Figure~\ref{fig:vel_rprof}. Inside the convective core (i.e., to the left of the dash-dotted vertical line), we can see the signature of the large-scale flow pattern described in Section~\ref{sec:renderings}. In particular, $|U_r|$ decreases as the flow nears the boundary and is deflected to travel mostly in the tangential direction, in turn explaining why $|U_t|$ increases in the same region. In the radiative envelope, the internal gravity waves leave an imprint on the $|U_t|$ profile, with a series of oscillations corresponding to the ring-like structures visible in the top-left panel of Figure~\ref{fig:bobs}. These features are similar to those observed in previous work on shell convection \citep{herwig2006,woodward2015,jones2017,andrassy2020,stephens2021} and core convection \citep{herwig2022}.

Figure~\ref{fig:vel_rprof} can also be used to assess the numerical convergence of our simulations. Separate runs using three different grid resolutions (up to $1728^3$) are shown, and all three use the same heating luminosity ($L=100 L_{\star}$). Clearly, the flow velocities in the convection zone change very little upon increasing the resolution, signalling a good convergence. In the envelope, high-wavenumber waves, which are less well resolved at lower grid resolutions, contain a significant amout of the total power (see Appendix~\ref{sec:kw}), and higher resolution translates into higher velocities. However, the velocity difference decreases when going from a $1152^3$ to a $1728^3$ grid compared to going from a $768^3$ to a $1152^3$ grid indicating that we are approaching convergence regarding the flow properties in the radiative envelope (which is not the main target of this investigation).

\subsection{Power spectra}
\label{sec:ps}
\begin{figure}
	\includegraphics[width=\columnwidth]{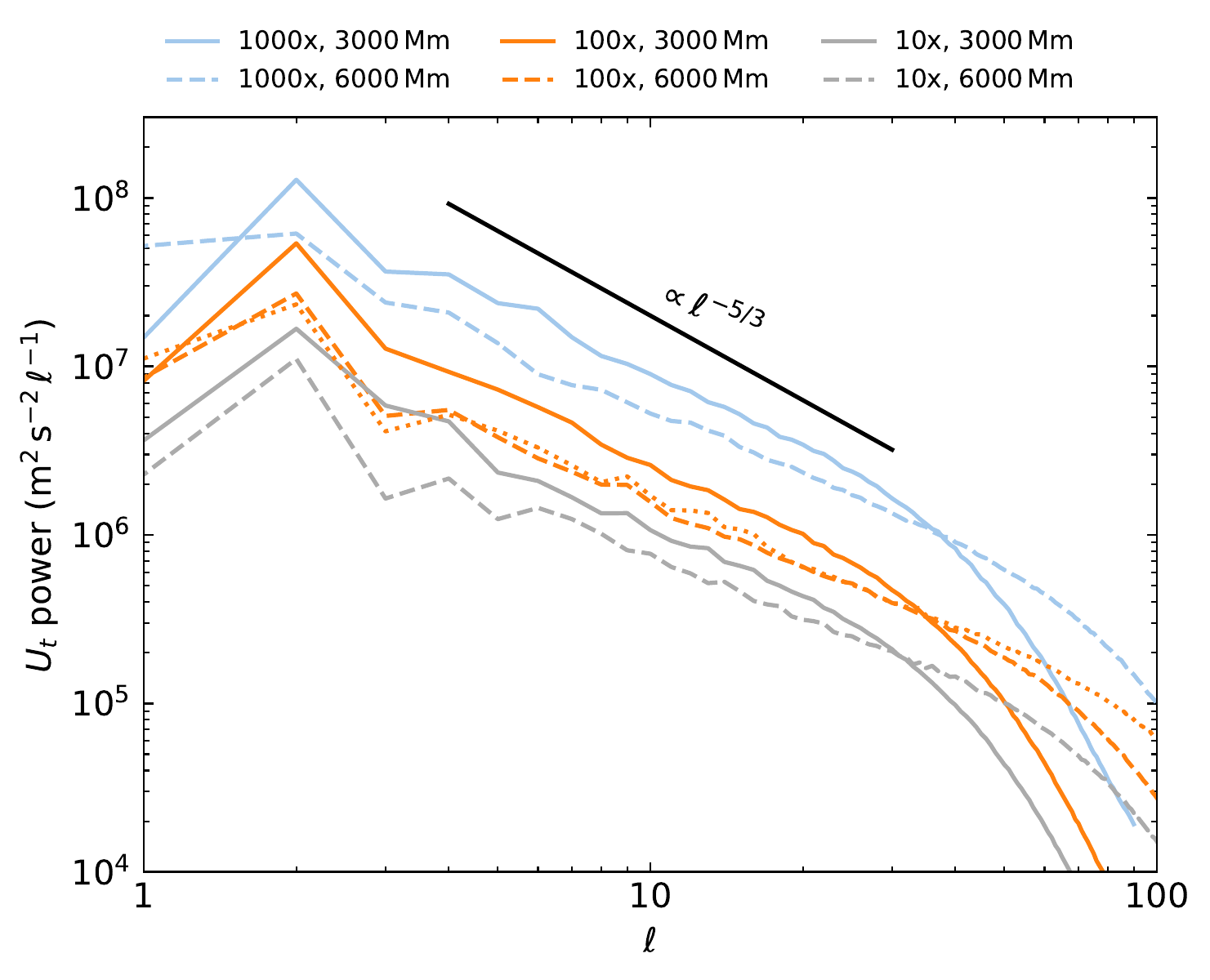}
	\includegraphics[width=\columnwidth]{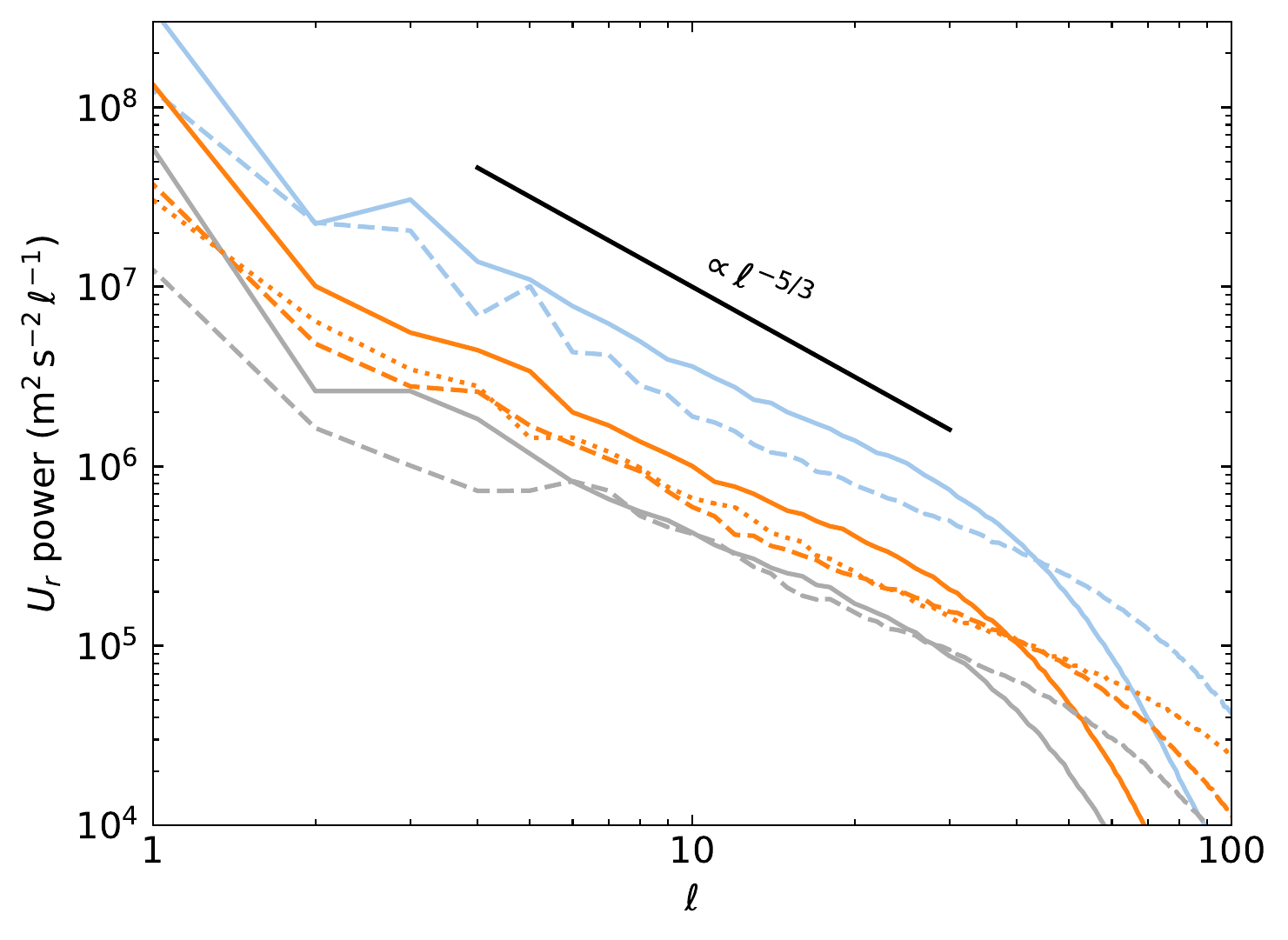}
    \caption{Power spectra of the tangential (top panel) and radial (bottom panel) velocity components for different heating rates and radii inside the convective core (see legend above the top panel). The power is binned as a function of $\ell$, the spherical harmonics angular degree. The spectra are averaged over the last 200 dumps of each simulation, and the solid and dashed lines correspond to simulations V3, V4 and V5 (1152$^3$ grid). An $\ell^{-5/3}$ Kolmogorov power law is shown for comparison. The dotted orange line shows the power spectrum for V7 (100x heating, $1728^3$ grid) at $R=6000\,{\rm Mm}$.}
    \label{fig:power}
\end{figure}

Now that we have examined the spherical averages of the velocity components, we turn to their fluctuations on the sphere at a given radius. To do so, we have decomposed the power contained in the flow into spherical harmonics (Figure~\ref{fig:power}). This is done using the filtered briquette data output \citep{stephens2021} for which the grid resolution in each direction is four times lower than the computational grid. For each velocity component, we show how much power is contained within a given spherical harmonics (identified by its spherical wavenumber $\ell$). We have repeated this exercise for two different radii inside the convective core. For $U_r$, note how the $\ell=1$ mode contains the most power, consistent with the large dipole structure visible in the top-right panel of Figure~\ref{fig:bobs}. Up to high $\ell$ ($\ell \lesssim 30$ at $R=6000\,{\rm Mm}$ and $\ell \lesssim 50$ at $R=3000\,{\rm Mm}$), we recover a Kolmogorov $\ell^{-5/3}$ power law, as expected for a turbulent convective flow. This demonstrates that the $P_{\rm rad}$-dominated nature of the equation of state has no influence on the smaller scale structure of the convective flow. Note that the departure from the Kolmogorov scaling at large $\ell$ simply reflects the finite grid resolution. With a higher resolution the $\ell^{-5/3}$ power law would continue to yet higher $\ell$. In fact, as revealed by a comparison of the $R=6000\,{\rm Mm}$ and $R=3000\,{\rm Mm}$ power spectra, the Kolmogorov scaling also extends to higher $\ell$ when the radius at which the power spectrum is calculated increases, since the angular resolution of the Cartesian simulation grid projected on the sphere is improved. A similar extension of the Kolmogorov scaling can also be observed for V7 (dotted orange line in Figure~\ref{fig:power}), which was performed on a 1728$^3$ grid instead of 1152$^3$ for the other simulations displayed here.

\subsection{Convective boundary}
\label{sec:boundary}
\begin{figure}
	\includegraphics[width=\columnwidth]{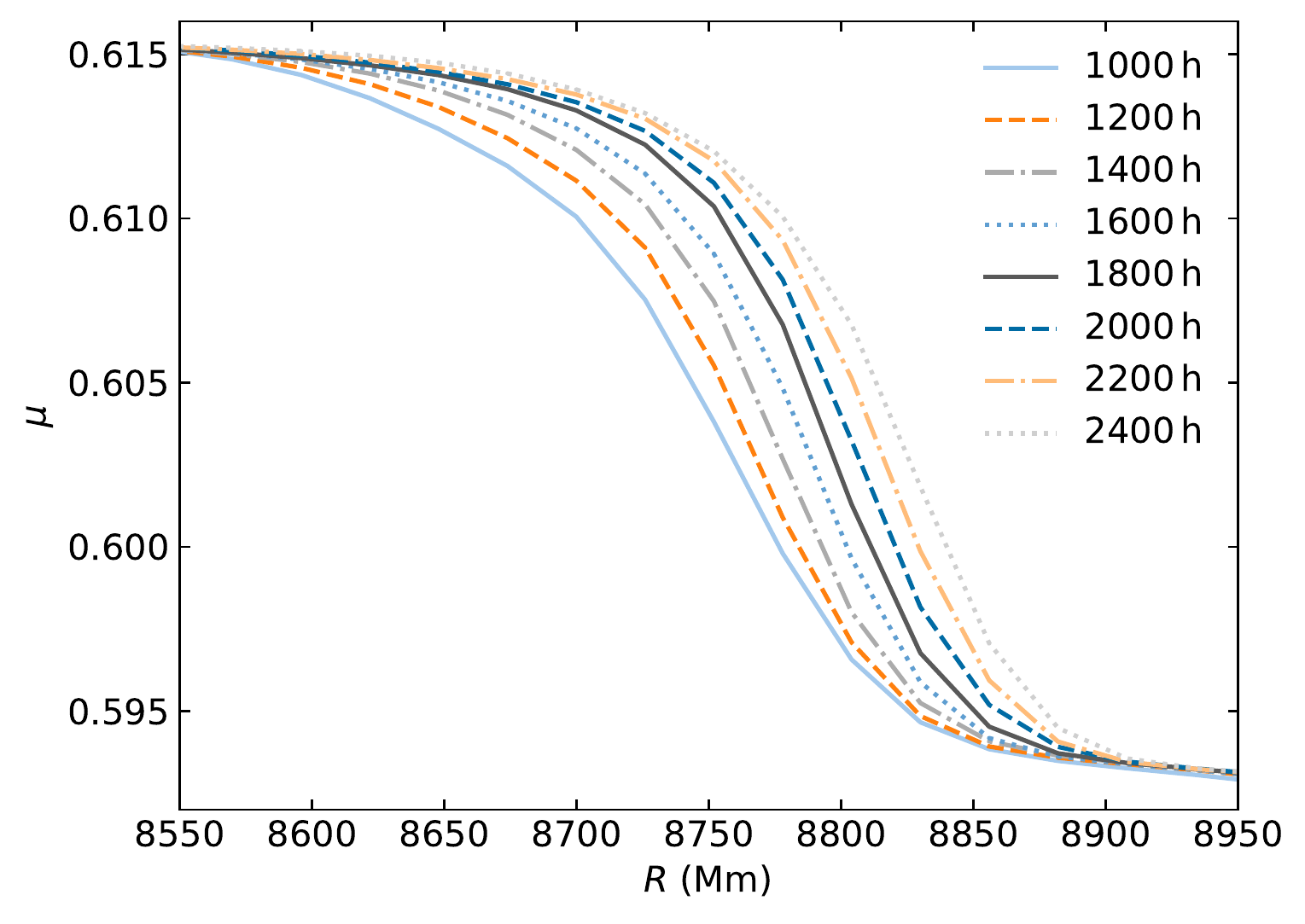}
    \caption{Evolution of the spherically averaged $\mu$ profile in run V3 ($L=1000 L_{\star}$ and $1152^3$ grid). A continuous outward migration of the convective boundary is observed, which we attribute to the fact that the initial setup is not thermally relaxed.}
    \label{fig:boundary}
\end{figure}

We have seen in Section~\ref{sec:simulations} how the properties of the flow in the convective core reach a steady state after a few thousand hours. This is to be contrasted with the behaviour of the convective boundary. Figure~\ref{fig:boundary} shows the evolution of the $\mu$ profile in the boundary region for run V3 ($L=1000 L_{\star}$ and $1152^3$ grid), which reveals that the convective boundary migrates outward at a rate of a few Mm per 100\,h. We attribute this behaviour to the fact that the 1D model used to initialize our 3D simulations is not thermally relaxed. This is not due to a flaw in our calculations but rather to a well-known feature of SMS: those stars never reach thermal equilibrium \citep{begelman2010}. The expansion of the convective core is simply an attempt by our 3D simulations to establish thermal equilibrium in the star. 

We cannot rule out that part of the migration of the convective boundary is due to genuine convective boundary mixing (penetration, overshoot, etc.). However, we cannot distinguish between this effect and the expansion due to the out-of-equilibrium nature of the initial setup, thereby preventing us from characterizing convective boundary mixing in SMS. In a future work, it may be interesting to generate 1D SMS models that are artificially relaxed to thermal equilibrium. 3D simulations initialized from such stratifications should have more stable convective boundaries, and it would then become possible to determine the properties of the boundary. Of course, the stratification would then differ from the true expected structure of rapidly-accreting SMS, but may provide an instructive experiment.

\subsection{Heating series}
\label{sec:heating_series}
\begin{figure}
	\includegraphics[width=\columnwidth]{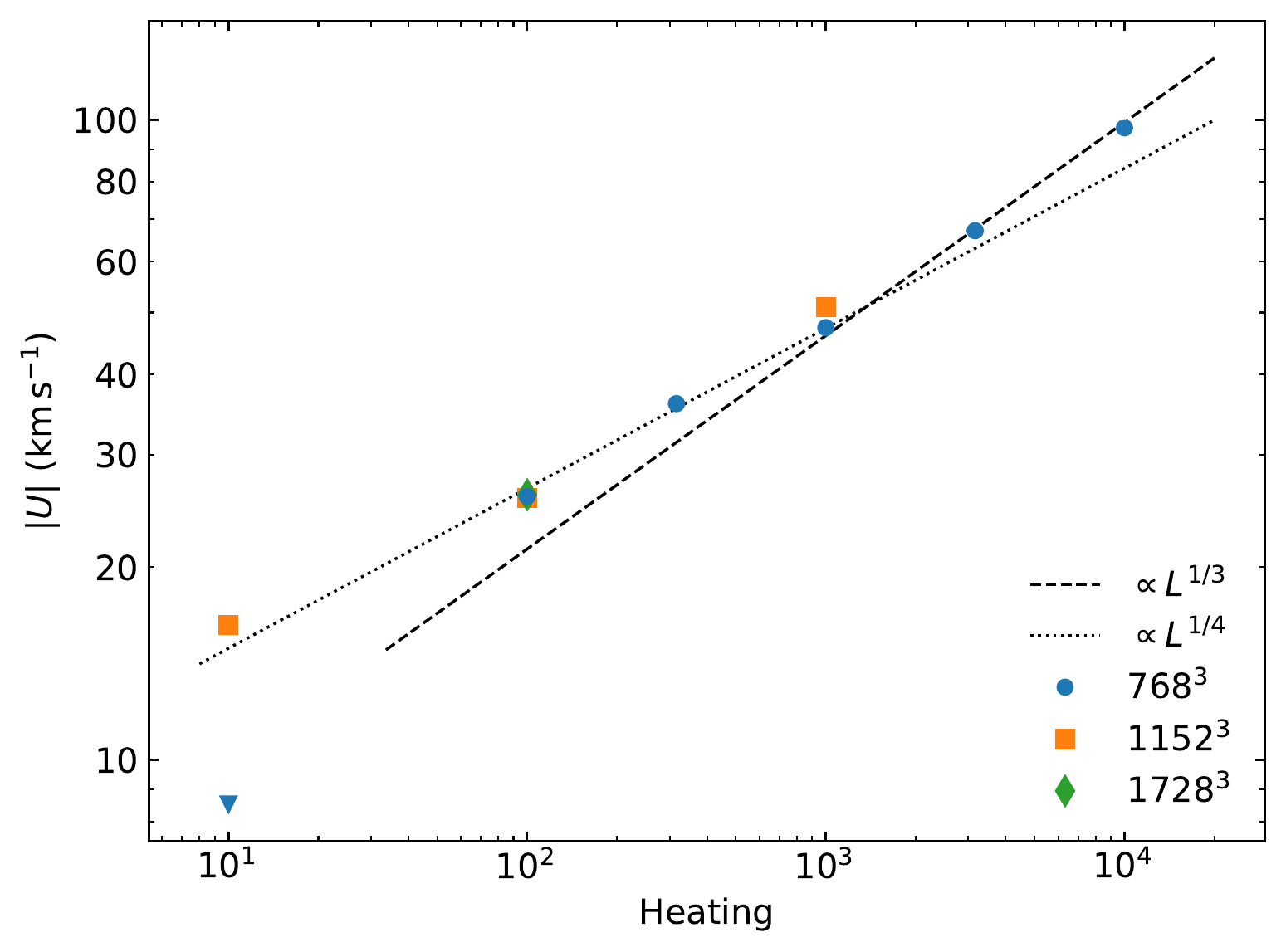}
    \caption{Spherically averaged rms velocity $1\,H_P (\simeq 2400\,{\rm Mm})$ below the convective boundary as a function of the driving luminosity. In each case, $|U|$ was averaged over the last 800\,h of the simulation; each run has reached dynamical equilibrium inside the convective core by that time. Note the superposition of the symbols for the three different grid resolutions in the $100\times$ heating case. The triangle symbol indicates that the underlying simulation is not fully converged (see text for details).}
    \label{fig:heating_series}
\end{figure}

In Figure~\ref{fig:heating_series}, we study the behaviour of the flow as a function of the heating luminosity by plotting the rms velocity one pressure scale height below the convective boundary for our ten simulations listed in Table~\ref{tab:runs}. Before delving into the luminosity dependence, let us examine how $|U|$ varies with respect to the grid resolution. At $100\times$ and $1000\times$ heating, we see that all grid resolutions agree very well with each other, as can be expected from our analysis of Figure~\ref{fig:vel_rprof}. In sharp contrast, the $768^3$ and $1152^3$ simulations at $L=10 L_{\star}$ strongly disagree, indicating poor numerical convergence. At this low heating rate, a $768^3$ grid is apparently too coarse to properly resolve the slow flow (${\rm Ma} \simeq 0.004$). In addition, despite running for more than 6000\,h, V12 (the $768^3$, $10 L_{\star}$ simulation) has not yet converged to a stable $|U|$ value. $|U|$ keeps decreasing, which is why we represent this simulation with a downward triangle in Figure~\ref{fig:heating_series}. For these reasons, we ignore this simulation in the following discussion.

Previous hydrodynamics simulations have found that the convective velocity scales as $L^{1/3}$ \citep[e.g.,][]{porter2000,muller2016,jones2017,herwig2022}. This is also what MLT predicts. Indeed, in the limit of large convective efficiency (which is appropriate here given the almost perfectly adiabatic core stratification), MLT gives \citep{cox1968}
\begin{equation}
    U = \frac{Q^{1/2}\alpha}{2 \sqrt{2} \Gamma_1^{1/2}} \left( \frac{\nabla_{\rm rad}-\nabla_{\rm ad}}{a_0 A} \right)^{1/3} c_s,
    \label{eq:v_mlt}
\end{equation}
where $c_s$ is the speed of sound, $\alpha$ is a free parameter of order unity, and $\nabla_{\rm rad}$ and $\nabla_{\rm ad}$ have their usual meanings. Adopting the proportionality constant $a_0=9/4$ of \cite{cox1968}, the dimensionless parameters of Equation~\eqref{eq:v_mlt} are given by
\begin{equation}
    A = \frac{Q^{1/2}c_p \kappa g \rho^{5/2} \ell_{\rm MLT}^2}{12 \sqrt{2} a c P^{1/2} T^3},
\end{equation}
\begin{equation}
    Q = \frac{4-3\beta}{\beta},
\end{equation}
and
\begin{equation}
    \Gamma_1 = \frac{32 - 24\beta - 3\beta^2}{24-21\beta},
\end{equation}
with $c_p$ the heat capacity at constant pressure and $\ell_{\rm MLT} \equiv \alpha H_P$. When we change the heating rate (and concurrently the opacity) in our simulations, only $A$ varies in Equation~\eqref{eq:v_mlt} because of the change in $\kappa$ ($\nabla_{\rm rad}$ remains constant because $L$ and $\kappa$ change by the same factor but in opposite directions). From there it follows that $U \propto L^{1/3}$ according to MLT, regardless of the value of $\beta$.\footnote{Interestingly, the same $L^{1/3}$ scaling is recovered if only the heating luminosity (and not the opacity) is changed. In that case, $\nabla_{\rm rad} = \frac{3}{16 \pi a c G} \frac{\kappa L P}{mT^4}$ increases linearly with $L$ while $A$ remains constant.}

Given these theoretical considerations and earlier 3D simulation results, it is perplexing to find in Figure~\ref{fig:heating_series} a luminosity scaling that disagrees with the expected $L^{1/3}$ scaling. At $L \geq 1000 L_{\star}$, our results are compatible with a $L^{1/3}$ power law, but at lower luminosities, they clearly favour a shallower dependence on $L$. This behaviour cannot be attributed to numerical convergence issues arising at low luminosities, since, as we have discussed, there is excellent agreement between different grid resolutions down to at least $L=100 L_{\star}$. Note that we reach the same conclusions if we study the scaling of the average rms velocity component in the whole convective core (instead of just looking at one particular radius), or if we examine the rms radial velocity component $|U_r|$ (instead of looking at $|U|$). We have not been able to identify a satisfying explanation for the unexpected $U-L$ relation revealed by Figure~\ref{fig:heating_series}. It is possible that this peculiar behaviour is related to the fact that the star is out of thermal equilibrium (see Section~\ref{sec:boundary}). Testing this hypothesis would require new simulations performed using an artificially thermally relaxed initial stratification.

\section{Mixing length theory analysis}
\label{sec:MLT}
In this section, we look in more details at simulation V3 ($L=1000 L_{\star}$ and $1152^3$ grid) and verify whether the properties of its convective core conform to predictions from MLT. Using other simulations for this analysis yields similar results, so we only focus on V3 for conciseness. The first thing we can compare is the superadiabadicity $\nabla-\nabla_{\rm ad}$. Since we are looking at the deep interior where convection is very efficient, the superadiabadicity is expected to be very small. In fact, it is so small that we cannot directly measure it in our simulation: the radial profile of $\nabla-\nabla_{\rm ad}$ oscillates around 0 with an amplitude close to the single-precision floating point precision of \code{PPMstar}. We can nevertheless constrain it to $\nabla-\nabla_{\rm ad} \lesssim 3\times 10^{-5}$ (measured $1\,H_P$ below the convective boundary), which is consistent with the MLT prediction of $3\times 10^{-6}$ (assuming $\alpha=1$).

\begin{figure}
	\includegraphics[width=\columnwidth]{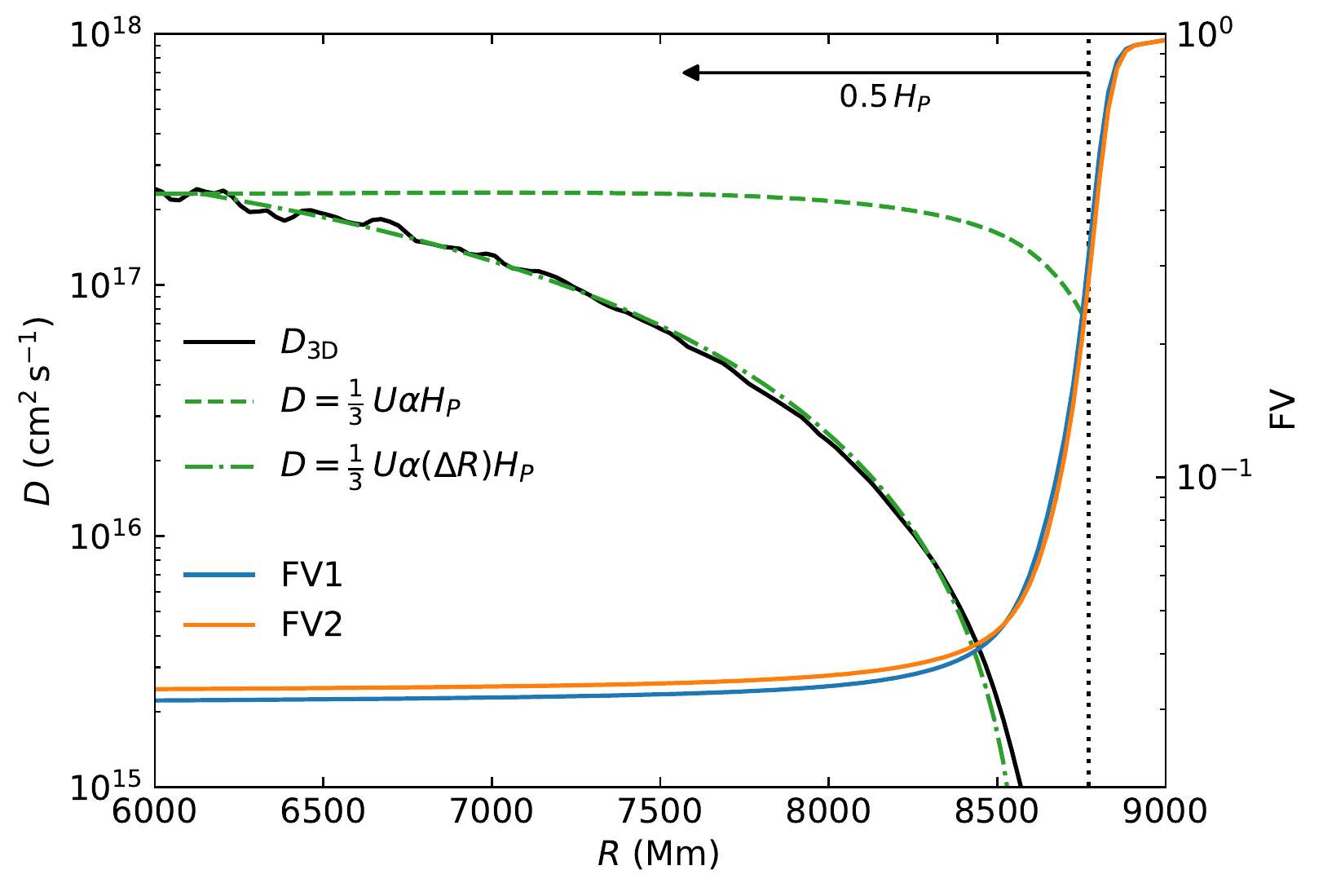}
    \caption{Diffusion coefficient (black solid curve) inferred from the change in the ${\rm FV}_{\rm env}$ profile (from ``FV1'' to ``FV2'') in the V3 simulation ($1000\times$ heating). The dashed line corresponds to the MLT formula $D = \frac{1}{3} U \alpha H_P$ with $\alpha=0.48$, while the dash-dotted line uses a non-constant $\alpha$ (Equation~\ref{eq:alpha_model}). The rms velocity $|U|$ profile from the 3D simulation was used to evaluate $D = \frac{1}{3} U \alpha H_P$. The vertical dotted line marks the location of the convective boundary.}
    \label{fig:diffusivity}
\end{figure}

Another quantity that we can compare to MLT is the diffusion coefficient in the convection zone. We measure this quantity in our simulation using the inversion method described in Section~2.4 of \cite{herwig2022}. Very briefly, we take ${\rm FV}_{\rm env}$ radial profiles at different times and invert the 1D diffusion equation to identify which diffusion profile $D(R)$ can reproduce the observed evolution. The result from this analysis is shown as a black solid line in Figure~\ref{fig:diffusivity}, where the ${\rm FV}_{\rm env}$ profiles used in the analysis are also shown for reference (blue and orange lines). Our inversion technique can only recover $D(R)$ if the gradient of ${\rm FV}_{\rm env}$ is not zero: this is why Figure~\ref{fig:diffusivity} is restricted to the outer portion of the convection zone.

Also shown in Figure~\ref{fig:diffusivity} is a dashed green line that corresponds to the prediction from the standard MLT formula $D=\frac{1}{3} U \alpha H_P$. We have assumed that $U$ in this equation corresponds to the rms velocity $|U|$ profile from the 3D simulation. The MLT diffusivity $1\,H_P$ inside the core matches the measured 3D value if we assume $\alpha=0.48$. However, the agreement closer to the boundary is poor. This is a well-known problem and a simple solution is to decrease the mixing length close to the boundary \citep{eggleton1972,jones2017}. This is what we did for the dashed-dotted line, where we now use the following prescription for $\alpha$:
\begin{equation}
\alpha(\Delta R) = \min \left(0.48, 0.54 \Delta R^2 + 0.19 \Delta R + 0.012 \right),
\label{eq:alpha_model}
\end{equation}
with
\begin{equation}
\Delta R = \frac{R_{\rm \scriptscriptstyle SB} - R}{H_P},
\end{equation}
where $R_{\rm \scriptscriptstyle SB}$ is the radius of the Schwarzschild boundary in the initial 1D model. Clearly, a much better agreement is now found. Note that a linear \citep{blouin2022} or exponential \citep{herwig2022} prescription for $\alpha(\Delta R)$ fails to replicates the measured diffusivity profile: a quadratic function provides a better fit. All things considered, the analysis presented in this section shows that MLT with $\alpha$ of order unity can reproduce the mixing measured in our simulations. Once again, this conclusion agrees with what has been previously established for stellar interior convective zones dominated by gas pressure.

\section{Conclusion}
\label{sec:conclusion}
We have performed the first 3D hydrodynamics simulations of core convection in primordial supermassive stars. We find that the peculiar conditions encountered in the interiors of those stars (in particular their radiation pressure-dominated nature) have no important effects on the properties of convection. We showed that the flow morphology is indistinguishable from that of core convection in massive main-sequence stars (where the gas pressure dominates), that the velocity spectra follow the expected Kolmogorov cascade, and that MLT with $\alpha$ of order unity can reliably describe mixing in the core. Our results offer compelling support for the use of MLT in 1D evolutionary models of supermassive stars.

Future work should focus on characterizing the properties of the convective boundary. This was not possible with our simulations as the convective boundary continually moves outward, a behaviour that we attribute to the fact that the star is not in thermal equilibrium. A customized 1D initial stratification where the star is allowed to relax to thermal equilibrium could conceivably be used for future 3D simulations aimed at measuring the properties of the convective boundary. Such simulations could also help elucidate the unexpected relation we have observed between the convective flow velocity and the heating luminosity.

\section*{Acknowledgements}
SB is a Banting Postdoctoral Fellow and a CITA National Fellow, supported by the Natural Sciences and Engineering Research Council of Canada (NSERC). TEW acknowledges support from the NRC-Canada Plaskett Fellowship. FH acknowledges funding through an NSERC Discovery Grant. PRW acknowledges funding through NSF grants 1814181 and 2032010. FH and PRW have been supported through NSF award PHY-1430152 (JINA Center for the Evolution of the Elements). The simulations presented in this work were carried out on the NSF Frontera supercomputer operated by the Texas Advanced Computing Center at the University of Texas at Austin and on The Alliance Niagara supercomputer operated by SciNet at the University of Toronto. The data analysis was carried on the Astrohub online virtual research environment (\url{https://astrohub.uvic.ca}) developed and operated by the Computational Stellar Astrophysics group (\url{http://csa.phys.uvic.ca}) at the University of Victoria and hosted on the Compute Canada Arbutus Cloud at the University of Victoria.

\section*{Data Availability}
Simulation outputs are available at \url{https://www.ppmstar.org} along with the Python notebooks that have been used to generate the figures presented in this work.

\bibliographystyle{mnras}
\bibliography{references}

\appendix
\section{Wavenumber--frequency diagrams}
\label{sec:kw}
\FloatBarrier

\begin{figure}
    \centering
\includegraphics[width=\columnwidth]{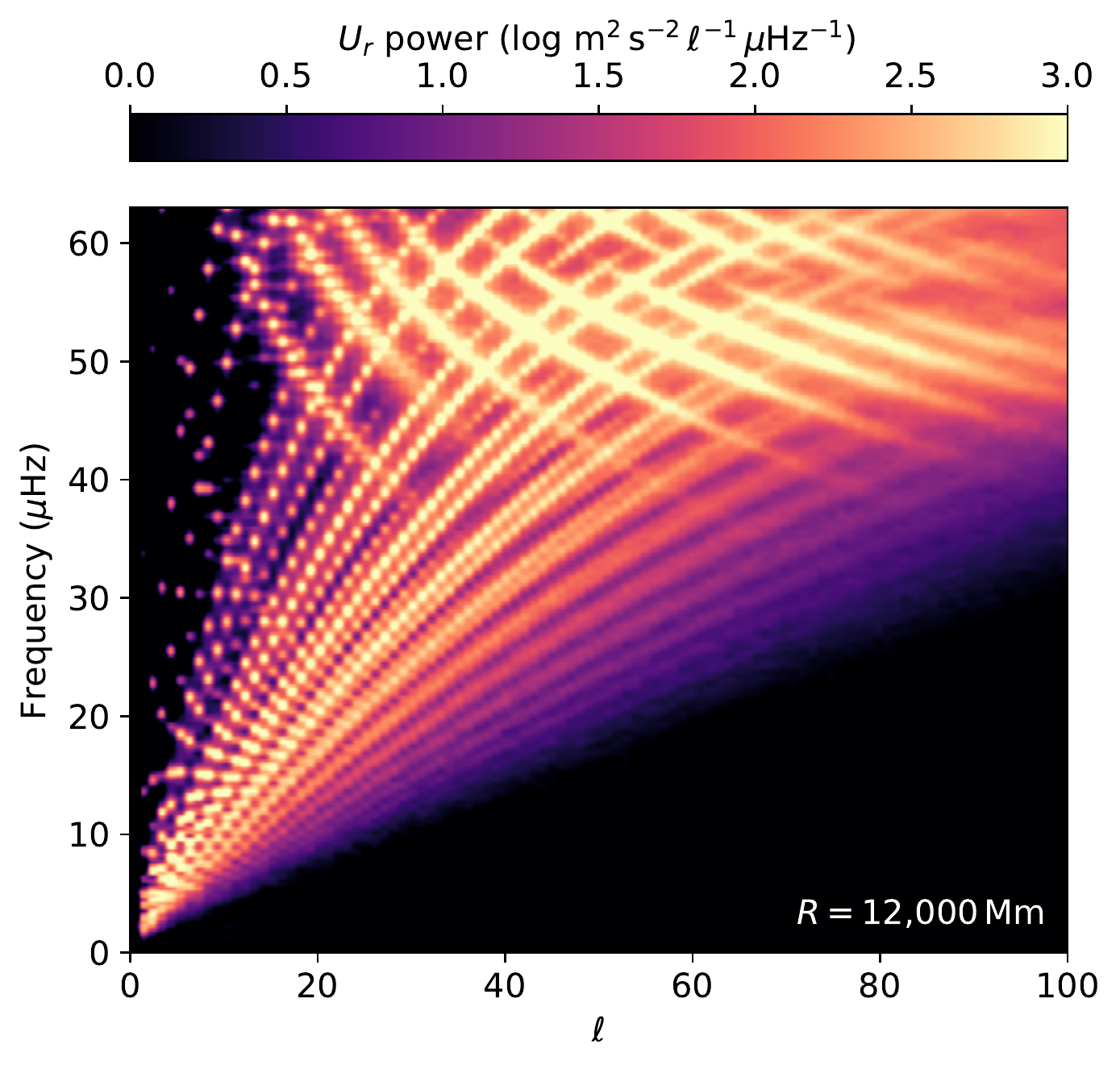}
    \includegraphics[width=\columnwidth]{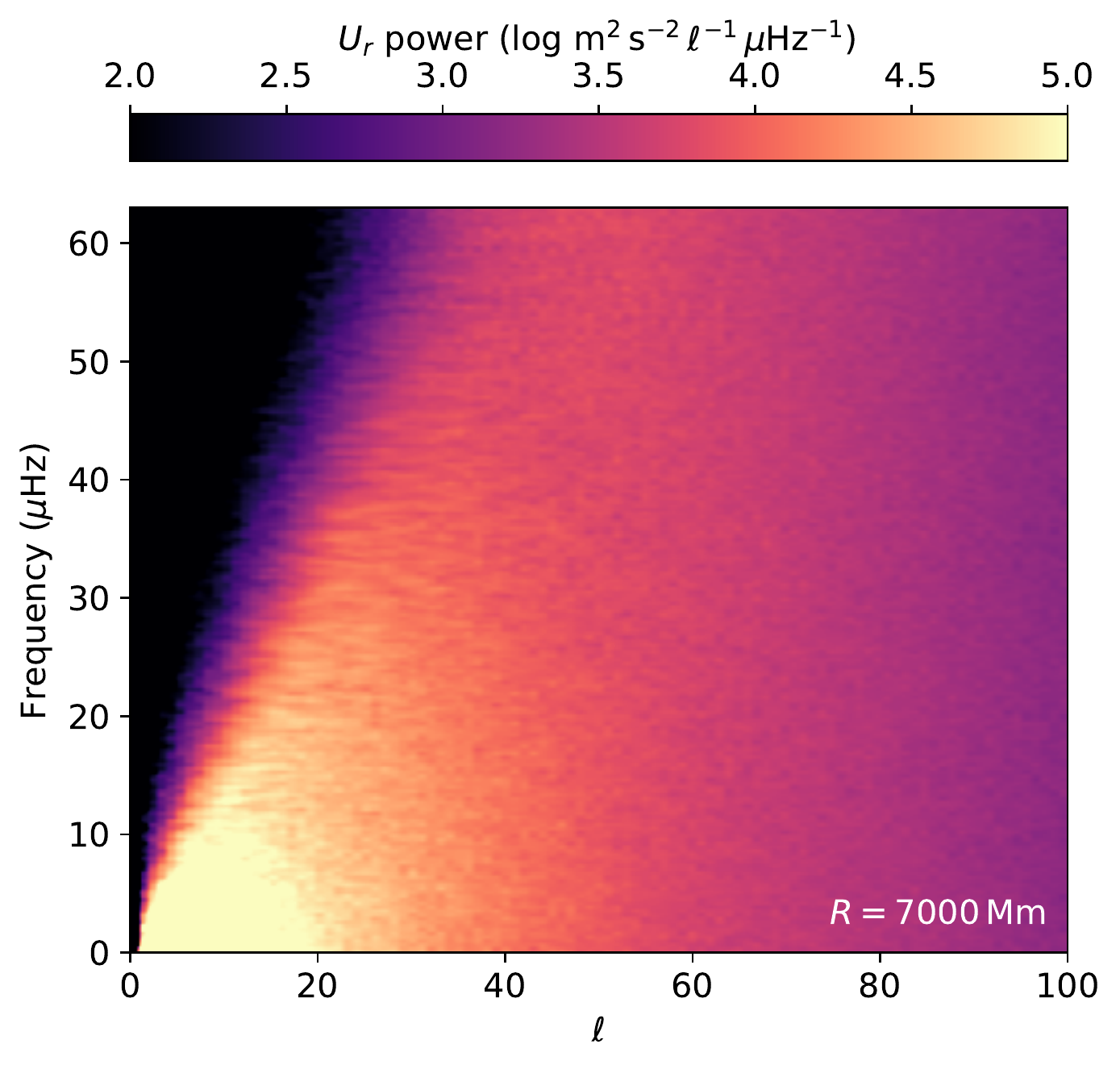}
    \caption{Power spectra of $U_r$ at $R=12{,}000\,$Mm (top panel) and $R=7000\,$Mm (bottom panel) as a function of the angular degree $\ell$ and the temporal frequency for V3 (1152$^3$ grid, $1000\times$ heating). The spectra were obtained by considering the last 400~dumps of the simulation. Note the different colour scale between the two panels.}
    \label{fig:kw}
\end{figure}

We show in Figure~\ref{fig:kw} power spectra of the radial velocity component for simulation V3 ($L=1000 L_{\star}$, $1152^3$ grid). The power is binned as a function of the angular degree (as in Section~\ref{sec:ps}) and of the temporal frequency. The wavenumber--frequency diagrams stop at $\nu=63\,\mu$Hz, which corresponds to the Nyquist cut-off frequency given the time interval that separates each detailed output of the simulation. Higher frequency modes are resolved in the simulation but cannot be reconstructed from the outputs (there are $\sim 2000$ simulation timesteps between each dump). The top panel shows the power spectrum in the stable layers (at $R=12{,}000\,$Mm), while the bottom panel shows the spectrum in the convective core (at $R=7000\,$Mm). In the convection zone, the spectrum is smooth and does not display any specific features, as expected for a turbulent flow. In contrast, in the radiative envelope, we see a distinctive power distribution, with a set of well-defined ridges composed of discrete modes. With increasing $\ell$, $\nu$ increases for most ridges but decreases for some. The first behaviour is the signature of internal gravity waves, also known as $g$ modes in asteroseismology \citep[compare for example Figure~\ref{fig:kw} to similar diagrams shown in][]{rogers2013,alvan2014,horst2020,thompson2022}. Internal gravity waves have frequencies below the local Brunt--V\"ais\"al\"a frequency, which here is $N/2\pi = 79\,\mu$Hz at $R=12{,}000\,$Mm. While we cannot resolve such high frequencies, the arcking of the ridges does suggest a convergence to the Brunt--V\"ais\"al\"a frequency at large $\ell$. As for the ridges that have decreasing frequencies with increasing $\ell$, they are artifacts of our Fourier decomposition. They correspond to aliases of internal gravity waves with $\nu>63\,\mu$Hz that are reflected at the Nyquist frequency.

\bsp
\label{lastpage}
\end{document}